\documentclass[preprintnumbers, floatfix,letterpaper,aps,prd,epsfig,nofootinbib,superscriptaddress,
twocolumn
]{revtex4-1}
\pdfoutput=1
\usepackage{bm,graphicx,dcolumn,epstopdf,epsf, latexsym,mathbbol, amssymb,amsmath,color,slashed, mathrsfs,mathcomp,simplewick}
\usepackage[center]{subfigure}
\usepackage{makecell}
\usepackage[colorlinks,linkcolor=blue,citecolor=blue,urlcolor=blue]{hyperref}

\begin{document}

\allowdisplaybreaks

\title{ Higgs Inflation with a Gauss-Bonnet term}

\author{Seoktae Koh}
\email{kundol.koh@jejunu.ac.kr}
\affiliation{Department of Science Education, Jeju National University, Jeju, 63243, Korea}
\affiliation{Institute for Gravitation and the Cosmos, Pennsylvania State University, University Park, PA 16802, USA}

\author{Seong Chan Park}
\email{sc.park@yonsei.ac.kr  (Corresponding author)}
\affiliation{Department of Physics \& IPAP \& Lab for Dark Universe, Yonsei University, Seoul 03722, Korea}
\affiliation{School of Physics, Korea Institute for Advanced Study, Seoul 02455, Korea}

\author{Gansukh Tumurtushaa}
\email{gansukh@jejunu.ac.kr (Corresponding author)}
\affiliation{Department of Science Education, Jeju National University, Jeju, 63243, Korea}

\begin{abstract}

Higgs inflation with a Gauss-Bonnet term is studied in the Einstein frame. Our model features two coupling functions, $\Omega^2(\phi)$ and $\omega(\phi)$, coupled to the Ricci scalar and Gauss-Bonnet combinations. We found a special relation $\Omega^2 \propto \omega$ sets the system a lot more simplified; therefore, we take it for granted in our analytical studies. 
As a result of a Weyl transformation to the Einstein frame, we notice the emergence of new interactions: a non-minimal kinetic coupling between the scalar field and gravity and a derivative self-interaction of the scalar field. In the Einstein frame, we investigate the cosmological implications of these interactions by deriving the background equation of motion and observable quantities. Our numerical result on $n_S$ vs. $r$ suggests our model is consistent with the observational data for a wide range of the model parameter, $-1.4\times 10^4\lesssim \alpha \equiv \frac{\omega}{\Omega^2} \lesssim 8\times 10^3$, where both the positive and negative values of $\alpha$ are allowed. As the Gauss-Bonnet contributions decay away with time after inflation, the propagation speed of gravitational waves turned out to be consistent with the recent constraints on the propagation speed of gravitational waves (GWs) without inducing ghost instability.

\end{abstract}
\maketitle

\section{Introduction}
\renewcommand{\theequation}{1.\arabic{equation}}\setcounter{equation}{0}

Cosmic inflation, an idea of accelerated exponential expansion of the early universe, is a successful paradigm that not only solved the flatness and horizon problems but also made definite predictions for primordial cosmological perturbations that observations can directly test; see Ref.~\cite{Sato:2015dga} for review. However, there is no conclusive solution to the problem of how to embed inflation into a particle physics framework. The most common approach for embedding inflation into the particle physics framework is to couple the gravity sector to a scalar field, such as the Higgs field. Driven by the Higgs field $\phi$, which is non-minimally coupled to gravity, Higgs inflation is a minimal model of inflation without introducing additional scalar degrees of freedom to those appearing in the Standard Model (SM) of particle physics~\cite{Bezrukov:2007ep, Bezrukov:2008ej, Bezrukov:2010jz, Bezrukov:2014ipa, Futamase:1987ua, Fakir:1990eg, Kamada:2012se}. This model agrees with data from Cosmic Microwave Background (CMB) experiments on the bounds of the scalar spectral index $n_S$ and the tensor-to-scalar ratio $r$~\cite{Komatsu:1999mt, Tsujikawa:2004my, Linde:2011nh, Planck:2013jfk, Tsujikawa:2013ila}. What makes Higgs inflation consistent with the observational data is the non-minimal coupling function between the Higgs field and the gravitational sector, which flattens the potential in the Einstein frame in the large-field regime, allowing the slow-roll conditions for inflation to be realized~\cite{Park:2008hz, Hyun:2022uzc, Hyun:2023bkf}.~\footnote{See \cite{Hyun:2022uzc, Hyun:2023bkf} for the non-minimal coupling of assistant field(s). } 
As a result, at first order in slow-roll approximation, the Higgs inflation model predicts a $n_S$ value consistent with data and a $r$ value to be comfortably below the experimental limits~\cite{Hamada:2014iga, Hamada:2014wna}; see \cite{Cheong:2021vdb} for the recent review. 

While the non-minimal coupling between gravity and the Higgs field is well-motivated by consideration of the renormalization of a scalar field in curved space, it is feasible to expect additional interactions to be present. 
From the effective field theory viewpoint $R^{2}$ term~\cite{Starobinsky:1979ty, Kannike:2015apa, Salvio:2015kka, Salvio:2016vxi, Calmet:2016fsr, Ema:2017rqn, He:2018gyf, Gorbunov:2018llf,  He:2018mgb, Canko:2019mud, Cheong:2019vzl,Cheong:2022gfc}, especially the Gauss-Bonnet combination $R_{GB}^2=R^2-4R_{\mu\nu}R^{\mu\nu}+R_{\mu\nu\rho\sigma}R^{\mu\nu\rho\sigma}$, are expected to arise~\cite{Weinberg:2008hq}. Higher curvature terms, $R^{2+p}$ terms (of mass dimension $4+2p$) may also arise, but they are supposed to be  suppressed~\cite{Cheong:2020rao, Jinno:2019und}. The Gauss-Bonnet term, in isolation, is purely topological and, therefore, does not impact the dynamics of inflation. However, it can introduce intriguing phenomenological effects when coupled with the inflaton field. Therefore, in the present work, we are motivated to study inflation in the context of a scalar field non-minimally coupled to the Ricci scalar and the Gauss-Bonnet combination. Such motivations for adding the Gauss-Bonnet term are also complemented by the string theory perspective, where particular couplings between the Gauss-Bonnet term and scalar fields have been found~\cite{Kawai:1998ab, Kawai:1999pw}. We note that many authors have studied phenomenological aspects of the Gauss-Bonnet combination, including cosmic inflation~\cite{Oikonomou:2020tct, Odintsov:2020mkz, Chakraborty:2018scm, Odintsov:2020xji, Oikonomou:2020oil, Pozdeeva:2020apf, Oikonomou:2020sij, Pozdeeva:2020shl, Odintsov:2019clh, Nojiri:2019dwl, Yi:2018gse, Satoh:2010ep, Guo:2009uk, Satoh:2008ck, Kawai:1998ab, Kawai:1999pw, Satoh:2007gn, Koh:2014bka}, primordial black holes~\cite{Kawai:2021edk}, gravitational-wave leptogenesis~\cite{Kawai:2017kqt}, dark energy~\cite{Nojiri:2005vv, Koivisto:2006xf, Nojiri:2007te, Amendola:2007ni, Granda:2014zea, Lee:2020upe, Bayarsaikhan:2022rzc, TerenteDiaz:2023iqk}, blackholes~\cite{Antoniou:2017acq, Lee:2018zym}, and wormholes~\cite{Tumurtushaa:2018agq, Chew:2020lkj}, in the Einstein frame version of a theory, where a generic function of a scalar field coupled to the Gauss-Bonnet combination is often considered in addition to the Einstein-Hilbert term. The Jordan frame analyses of Higgs inflation and a primordial black hole with the Gauss-Bonnet term, on the other hand, were investigated in Refs.~\cite{vandeBruck:2015gjd, Kawaguchi:2022nku}, respectively.

The paper is organized as follows. Section~\ref{sec:2} begins with our setup formulated in the Jordan frame, where we have a scalar (or Higgs) field coupled to the Ricci scalar and the Gauss-Bonnet combination. At the end of the section, we obtain the Einstein frame action using the so-called conformal transformation. From the Einstein frame action, we derive the background equations of motion and the observable quantities in Section~\ref{sec:3} following Ref.~\cite{Hwang:2005hb}. In the same section, we provide our numerical results and discuss the consequent findings of our work. Finally, we conclude our work in Section~\ref{sec: conclusion}.

\section{Setup and Conformal transformation} \label{sec:2}
\renewcommand{\theequation}{2.\arabic{equation}}\setcounter{equation}{0}
Let us begin with an action given by
\begin{align}\label{eq: action}
    S^J&=\int d^4x \sqrt{-g^J}\left[ 
    \frac{M_p^2}{2} \Omega^2(\phi) R^J \right. \nonumber\\
    &\qquad \left.- \frac{1}{2}g^J_{ab}\nabla^a\phi\nabla^b\phi-V(\phi)+\omega(\phi)R_{GB}^{2^J}\right]\,,
\end{align}
where $M_p$ is the reduced Planck mass. The superscript $J$ denotes quantities in the Jordan frame, where the scalar field $\phi$ is coupled to the Ricci scalar $R$ of the gravity sector through the non-minimal coupling function $\Omega(\phi)$. The $\omega(\phi)$ is the coupling function between the $\phi$ and the Gauss-Bonnet combination, $R_{GB}^2=R^2-4R_{\mu\nu}R^{\mu\nu}+R_{\mu\nu\rho\sigma}R^{\mu\nu\rho\sigma}$. If one interprets the $\phi$ field as the unitary-gauge Higgs field, the first three terms in Eq.~(\ref{eq: action}) are well known in the context of Higgs inflation, for which the non-minimal coupling function $\Omega^2(\phi)$ and the potential $V(\phi)$ take the following forms~\cite{Bezrukov:2007ep, Bezrukov:2010jz, Bezrukov:2008ej, Futamase:1987ua, Fakir:1990eg}
\begin{align}\label{eq: higgspot}
    \Omega^2 = 1+\frac{\sigma}{M_p^2}\phi^2\,, \quad V(\phi)=\frac{\lambda}{4} \left(\phi^2-v^2\right)^2\,,
\end{align}
where $\sigma$ and $\lambda$ are the coupling constant and the potential parameters, respectively. The $v$ is the vacuum expectation value of the Higgs field, i.e., $v\sim \mathcal{O}(10^2)\,$GeV, which can be neglected at large field limit ($\phi\gg v$). Thus, the quartic potential $V(\phi)\simeq \lambda \phi^4/4$ is a good approximation during Higgs inflation. 

Many properties of the physically interesting quantities become more apparent and easier to present in the Einstein frame, where the scalar field is minimally coupled to the Ricci scalar of a gravity sector. Using the so-called Weyl transformation, a local conformal transformation, one moves from the Jordan frame to the Einstein frame. The spacetime metric and the square root of its determinants change under the conformal transformation as
\begin{align}\label{eq: conformalmetric}
    g_{ab}^J = \Omega^{-2} g_{ab}\,, \quad \sqrt{-g^J}=\Omega^{-4}\sqrt{-g},
\end{align}
where the $g_{ab}$, without the superscript $J$, is the metric in the Einstein frame. In the Einstein frame, the action is written as~\cite{Bezrukov:2007ep, Bezrukov:2010jz, Bezrukov:2008ej, Futamase:1987ua, Fakir:1990eg}
\begin{align}\label{eq: actioninE}
    S=\int d^4 x \sqrt{-g}\left[\frac{M_p^2}{2}R -\frac12 g_{ab}\nabla^a s\nabla^b s -V(s) \right]\,,
\end{align}
where $s$ is the new canonical scalar field, which is related to the $\phi$ via
\begin{align}\label{eq: relation}
    \frac{ds}{d\phi} =\sqrt{\frac{1}{\Omega^2}+\frac{3M_p^2}{2}\left(\frac{d\ln\Omega^2}{d\phi}\right)^2}\nonumber\\ =\left[ \frac{1+\sigma\left(1+6\sigma\right)\phi^2/M_p^2}{\left( 1+\sigma\phi^2/M_p^2\right)^2}\right]^{1/2}
    \,,
\end{align}
and $V(s)\equiv V(\phi(s))/\Omega^{4}(\phi(s))$ is the Einstein frame potential. Eq.~(\ref{eq: relation}) can be solved for $s(\phi)$ as
\begin{align}\label{eq: canonicals}
    \frac{s}{M_p} =\sqrt{\frac{1+6\sigma}{\sigma}}\,\text{arcsinh}\left[\sqrt{\sigma(1+6\sigma)}\phi/M_p\right] \nonumber \\
    -\sqrt{6}\,\text{arctanh}\left[
    \frac{\sqrt{6}\sigma \phi/M_p}{\sqrt{1+\sigma(1+6\sigma)\phi^2/M_p^2}}
    \right]\,.
\end{align}
In the large coupling limit $\sigma|\phi|/M_p\gg 1$ limit, Eq.~(\ref{eq: higgspot}) and Eq.~(\ref{eq: canonicals}) can be well approximated as~\cite{Cheong:2021vdb, He:2018mgb}
\begin{align}\label{eq:sfield}
    \frac{s}{M_p}\simeq \sqrt{\frac32}\ln\Omega^2(\phi(s))\,.\quad 
\end{align} 
Thus, by substituting this into the potential, we get
\begin{align}\label{eq: einsteinpot}
    V(s) \simeq \frac{\lambda M_p^4}{4 \sigma^2}\left(1-e^{-\sqrt{\frac23}\frac{s}{M_p}}\right)^2\,.
\end{align}
Let us now discuss how the last term in Eq.~(\ref{eq: action}) transforms under the conformal transformation and investigate what consequent dynamics would be apparent in the Einstein frame that otherwise does not come into sight in the Jordan frame. 

The last term of the action in Eq.~(\ref{eq: action}) reads
\begin{align}\label{eq: actionGB}
    S^J_{GB}=\int d^4x \sqrt{-g^J} \omega(\phi)R_{GB}^{2^J}\,.
\end{align}
The Gauss-Bonnet combination changes under the conformal transformation as~\cite{Carneiro:2004rt}
\begin{align}\label{eq: jordanGB}
    R_{GB}^{2^J} &= \Omega^4\left[R_{GB}^2-8\Omega^{-1}G_{ab}\nabla^a\nabla^b\Omega
    -4R\Omega^{-2}\nabla_a\Omega\nabla^a\Omega \right. \nonumber\\
    &+8\Omega^{-2}\left(\nabla_a\nabla^a\Omega\nabla_b\nabla^b\Omega-\nabla_b\nabla_a\Omega\nabla^b\nabla^a\Omega \right) \nonumber\\  & \left.-24\Omega^{-3}\nabla_a\Omega\nabla^a\Omega\nabla_b\nabla^b\Omega +24\Omega^{-4}\left( \nabla_a\Omega \nabla^a\Omega\right)^2\right]\,,
\end{align}
where $G_{ab}\equiv R_{ab}-g_{ab}R/2$ is the Einstein tensor.
Substituting Eq.~(\ref{eq: jordanGB}) into Eq.~(\ref{eq: actionGB}) and using Eq.~(\ref{eq: conformalmetric}), we obtain the action in the Einstein frame as
\begin{align}\label{eq: actionGB1}
    S_{GB}&=\int d^4x\sqrt{-g}\, \omega(\phi)\nonumber \\
    &\times \left[R_{GB}^2-8\Omega^{-1}G_{ab}\nabla^a\nabla^b\Omega -4R\Omega^{-2}\nabla_a\Omega\nabla^a\Omega\right. \nonumber\\ 
    &+8\Omega^{-2}\left(\nabla_a\nabla^a\Omega\nabla_b\nabla^b\Omega-\nabla_b\nabla_a\Omega\nabla^b\nabla^a\Omega \right) \nonumber\\ &\left.-24\Omega^{-3}\nabla_a\Omega\nabla^a\Omega\nabla_b\nabla^b\Omega +24\Omega^{-4}\left( \nabla_a\Omega \nabla^a\Omega\right)^2\right]\,.
\end{align}
The coupling functions $\omega(\phi)$ can generally be either a constant or a generic function of a scalar field. In the Appendix~\ref{sec: appA}, we show that, if $\omega=\text{const.}$, no accountable effect comes from the Gauss-Bonnet term in both frames. Thus, from now on, we regard the Gauss-Bonnet coupling as a generic function of the scalar field. With the use of the integration by parts, the second and the third terms in Eq.~(\ref{eq: actionGB1}) can be simplified as
\begin{align}\label{eq: part1}
    -8\int d^4x\sqrt{-g}\left[\omega \Omega^{-2} R_{ab} \nabla^a\Omega\nabla^b\Omega -\Omega^{-1}G_{ab}\nabla^a\omega\nabla^b\Omega\right]\,,
\end{align}
while the fourth term becomes
\begin{align}\label{eq: part2}
    &8\int d^4x\sqrt{-g} \left[\omega\Omega^{-2} R_{ab} \nabla^a\Omega\nabla^b\Omega \right.\nonumber \\
    &\quad -\omega\Omega^{-2}\left(\omega^{-1}\nabla_a\omega-2\Omega^{-1}\nabla_a\Omega\right) \nonumber \\
    &\qquad\left.\times\left(\nabla^a\Omega\nabla_b\nabla^b\Omega -\nabla_b\Omega\nabla^a\nabla^b\Omega\right)\right]\,,
\end{align}
where we used Eq.~(\ref{eq: identity}). 
Consequently, Eq.~(\ref{eq: actionGB1}) can be rewritten as 
\begin{align}\label{eq: actionGB3}
    S_{GB}&=\int d^4x\sqrt{-g}\left[\omega R_{GB}^2 + 8\Omega^{-1}G_{ab}\nabla^a\omega\nabla^b\Omega\right.  \nonumber\\ &\quad-8\omega\Omega^{-2}\left(\omega^{-1}\nabla_a\omega-2\Omega^{-1}\nabla_a\Omega\right)\nonumber\\
    &\qquad\times\left(\nabla^a\Omega\nabla_b\nabla^b\Omega -\nabla_b\Omega\nabla^a\nabla^b\Omega\right) \nonumber\\ 
    &\qquad \quad -24\omega \Omega^{-3}\nabla_a\Omega\nabla^a\Omega\nabla_b\nabla^b\Omega \nonumber\\
    &\qquad\qquad\left.+24\omega\Omega^{-4}\left( \nabla_a\Omega \nabla^a\Omega\right)^2\right]\,,
\end{align}
where the first term in Eq.~(\ref{eq: part1}) is canceled with that of Eq.~(\ref{eq: part2}). It is worth noting that the third term in Eq.~(\ref{eq: actionGB3}) vanishes when the two non-minimal couplings are proportional to each other, maintaining the following relation:
\begin{align}
\omega =\alpha \Omega^2\,, \label{eq: omegas}
\end{align}
where $\alpha \in \mathbb{R}$. Although the coupling functions $\Omega^2(\phi)$ and $\omega(\phi)$ have the flexibility to be arbitrary functions of a scalar field, we would assume Eq.~(\ref{eq: omegas}) as a part of our model. The physical meaning behind this particular relation is that, with this choice, the Gauss-Bonnet term can be regarded as the next-to-leading order correction to the gravitational sector with coupling constant $\alpha$ as is seen in Eq.~(\ref{eq: AppB1}). Moreover, once this relation is granted, which we did in this work, the form of the Gauss-Bonnet coupling function can also be determined, such that our flexibility in choosing $\omega(\phi)$ is quite restricted. For further elaboration, including the form of action and the case of an arbitrary power relationship $\omega\propto \Omega^p$, please refer to Appendix~\ref{sec: appB}. 

Now, the action is greatly simplified as
\begin{align}\label{eq: actionGB4}
    S_{GB} &=\int d^4x\sqrt{-g}\,\alpha\Omega^2 \left[R_{GB}^2 + 4 G_{ab}\nabla^a \ln\Omega^2\nabla^b\ln\Omega^2 \right.\nonumber \\ &\quad \left.-3\nabla^b\nabla_b\ln\Omega^2\nabla_a\ln\Omega^2\nabla^a\ln\Omega^2\right]\,.
\end{align} 
In terms of the scalar field $s$ defined in Eq.~(\ref{eq:sfield}), the action in Eq.~(\ref{eq: actionGB4}) becomes
\begin{align}\label{eq: actionGB5}
    S_{GB}&=\int d^4x\sqrt{-g} \alpha e^{\sqrt{\frac{2}{3}}\frac{s}{M_p}}\left[R_{GB}^2 +\frac{8}{3M_p^2}G_{ab}\nabla^as \nabla^bs \right.\nonumber\\
    &\qquad\left. -\frac{1}{M_p^3}\sqrt{\frac{8}{3}}\nabla^b\nabla_bs\nabla_as\nabla^as\right]\,.
\end{align}
Combining Eq.~(\ref{eq: actionGB5}) with Eq.~(\ref{eq: actioninE}), we can write the full action in the Einstein frame as 
\begin{align}\label{eq: fullaction}
    &S=\int d^4 x \sqrt{-g}\left[\frac{M_P^2}{2}R -\frac12 g_{ab}\nabla^a s\nabla^b s- V(s)-\frac{1}{2}\xi(s)  \right.\nonumber\\ 
    &\times\left.\left(c_1 R_{GB}^2 +\frac{c_2}{M_p^2}G_{ab}\nabla^as \nabla^bs +\frac{c_3}{M_p^3}\nabla_as\nabla^as \nabla^b\nabla_bs\right)\right]\,,
\end{align}
where
\begin{align*}
    \xi(s)\equiv -2\alpha e^{\sqrt{\frac{2}{3}}\frac{s}{M_p}}\,,\quad c_1=1\,,\quad c_2=\frac{8}{3}\,,\quad c_3=-\sqrt{\frac{8}{3}}\,,
\end{align*}
and the potential $V(s)$ given in Eq.~(\ref{eq: einsteinpot}). Here, in addition to the expected Gauss-Bonnet term, Eq.~(\ref{eq: fullaction}) presents new interactions; including the kinetic coupling between the scalar field and gravity, as well as the derivative self-interaction of the scalar field, that were not apparent in the Jordan frame. Such interactions discussed in Ref.~\cite{Tumurtushaa:2019bmc, Bayarsaikhan:2020jww, Chien:2021zle} as a particular subclass of Horndeski's theory~\cite{Horndeski:1974wa}, the most general scalar-tensor theory of gravity, or equivalently the generalized Galileons~\cite{Kobayashi:2011nu}. Moreover, multiplied by $\xi(s)$, the last term in Eq.~(\ref{eq: fullaction}) is also discussed in Ref.~\cite{Hwang:2005hb} as a string correction to Einstein gravity. These additional interactions are dropped out if the $\alpha$ equals zero, such that the general relativity (GR) limit can be reached in our study. Thus, the deviation from the GR requires the non-zero values of $\alpha$; the larger the $\alpha$ value, the more the deviation from the GR limit increases. Consequently, even though both negative and positive values are allowed, we let the observational data determine the sign of $\alpha$. From now on, we will specify the last term of Eq.~(\ref{eq: fullaction}) as the Gauss-Bonnet contributions.

\section{Higgs inflation with a Gauss-Bonnet term in the Einstein frame}\label{sec:3}
\renewcommand{\theequation}{3.\arabic{equation}}\setcounter{equation}{0}

In this section, we investigate Higgs inflation with the Gauss-Bonnet contributions in the Einstein frame with potential presented in Eq.~(\ref{eq: einsteinpot}). From Eq.~(\ref{eq: fullaction}), we derive gravitational and field equations of motion as~\cite{Hwang:2005hb} 
\begin{subequations}\label{eq:EOM}
\begin{align}
&G_{ab} = \nabla_a s \nabla_b s - \frac{1}{2}g_{ab}\left(\nabla_c s \nabla^c s -2V\right)-\frac{1}{2}T_{ab}^{GB}\,, \\
&\nabla_a\nabla^a s  -V_{,s} = \frac{1}{2} T^{GB}\,,
\end{align}
\end{subequations}
where
\begin{widetext}
    \begin{align*}
    T_{ab}^{GB}&= c_1\left[4g_{ab}\left(2\nabla_{c}\nabla_d\xi R^{cd}-\nabla_c\nabla^c\xi R \right) -4\left(2 \nabla^{c}\nabla^d\xi R_{acbd}-2\nabla_c\nabla^cR_{ab}+4\nabla_{c}\nabla_{(b} \xi R_{a)}^c -\nabla_a\nabla_b \xi R\right)\right]\nonumber\\
    &\quad+\frac{c_2}{M_p^2}\left[\xi(R_{ab}\nabla ^c s \nabla_c s+R \nabla_a s \nabla_b s -4R^c_{(a} \nabla_{b)}s \nabla_{c}s)-\nabla_c\nabla^c\left(\xi \nabla_{a}s \nabla_{b}s \right) \right. \nonumber\\
    &\qquad\left.-\nabla_a\nabla_b\left( \xi \nabla^{c}s \nabla_{c}s\right)+2\nabla_c\nabla_{(b}(\xi \nabla^{c}s \nabla_{a)}s) +g_{ab} \left( \xi G^{cd}\nabla_{c}s\nabla_{d}s - \nabla_d \nabla_{c}(\xi \nabla^{c}s \nabla^{d}s) +\nabla_d\nabla^d(\xi \nabla_{c}s \nabla^{c}s)\right)\right] \nonumber\\
    &\quad +\frac{c_3}{M_p^3} \left[2\nabla_{(a}(\xi \nabla^{c}s \nabla_{c}s)\nabla_{b)}s-2\xi\nabla^c\nabla_c s\nabla_{a}s\nabla_{b}s - g_{ab}\nabla_{d}(\xi \nabla^{c}s\nabla_{c}s)\nabla^{d}s \right]\,,  \\
    T^{GB} &= c_1 \xi_{,s}R^2_{GB} - \frac{c_2}{M_p^2}G^{ab}\left(\xi_{,s}\nabla_{a}s\nabla_{b}s+2\xi \nabla_a\nabla_b s\right)\nonumber\\
    &\quad+ \frac{c_3}{M_p^3} \left[\xi_{,s}\nabla_b\nabla^b s \nabla^{a}s\nabla_{a}s + \nabla_b\nabla^b\left(\xi \nabla_{a}s\nabla^{a}s\right)-2\nabla_{a}\left(\xi\nabla_b\nabla^b s \nabla^{a}s\right) \right]\,,
    \end{align*}
\end{widetext}
with ``$V_{,s}=\partial V/\partial s$'' and ``$\xi_{,s}=\partial \xi/\partial s$.'' 
In the spatially flat FRW universe with metric 
\begin{align*}
    ds^2 = -dt^2+a(t)^2\delta_{ij}dx^idx^j\,,
\end{align*}
where $a(t)$ is the scale factor,
the background equations of motion yield from Eq.~(\ref{eq:EOM})~\cite{Hwang:2005hb}
\begin{subequations}\label{eq: bEOMs}
        \begin{align}\label{eq: bEOM1}
        &3M_p^2 H^2 = \frac{1}{2}\dot{s}^2 + V+ 12c_1\dot{\xi}H^3 - \frac{9}{2}\frac{c_2}{M_p^2} \xi \dot{s}^2H^2 \nonumber\\
        & \quad+\frac{1}{2}\frac{c_3}{M_p^3}(\dot{\xi}-6\xi H)\dot{s}^3\,,\\
        &M_p^2(2\dot{H}+3H^2) = \nonumber \\
        &-\frac{1}{2}\dot{s}^2 +V +4c_1\left[\ddot{\xi}H^2+2\dot{\xi}H(\dot{H}+H^2)\right] \nonumber\\
        &\quad -\frac{1}{2}\frac{c_2}{M_p^2}\dot{s}\left[\xi\dot{s}(2\dot{H}+3H^2)+4\xi\ddot{s}H +2\dot{\xi}\dot{s}H\right] \nonumber \\
        &\qquad-\frac{1}{2}\frac{c_3}{M_p^3}\dot{s}^2(2\xi\ddot{s}+\dot{\xi}\dot{s})\,,\label{eq: bEOM2}\\
        &\ddot{s}+3H\dot{s}+V_{,s}=-12c_1\xi_{,s}H^2(\dot{H}+H^2) \nonumber\\
        &\quad +\frac{3}{2}\frac{c_2}{M_p^2}\left[H^2(\dot{\xi}\dot{s}+2\xi \ddot{s})+2H\xi\dot{s}(2\dot{H}+3H^2)\right]\nonumber\\
        &\qquad -\frac{1}{2}\frac{c_3}{M_p^3}\dot{s}\left[ \ddot{\xi}\dot{s}+3\dot{\xi}\ddot{s} -6\xi(\dot{H}\dot{s}+2H\ddot{s}+3H^2\dot{s})\right]\,,\label{eq: bEOM3}
        \end{align}
    \end{subequations}
where $H\equiv \dot{a}/a$ is the Hubble parameter and the over-dot denotes the derivative with respect to time $t$. 
The imprints of the Gauss-Bonnet contributions in the Einstein frame can be easily identified by examining the equation of motion for the presence of the $\xi(s)$ function. Thus, the terms containing $\xi(s)$ are clear indicators of the Gauss-Bonnet contributions in the Einstein frame and should not be overlooked. 

In the context of slow-roll inflation, it is often assumed that the acceleration of the scalar field is negligible with respect to the gravitational friction, $\ddot{s}\ll 3H\dot{s}$, and the potential energy dominates over the kinetic energy, $V\gg\dot{s}^2/2$; together they are known as the slow-roll approximations. Thus, in light of the slow-roll approximations, the above equations in Eqs.~(\ref{eq: bEOMs}) can be simplified even further as
\begin{align}
    &3M_p^2H^2\simeq V\,,\label{eq:friedman}\\
    &3H\dot{s} \simeq -\frac{\mathcal{B}\mp \sqrt{\mathcal{B}^2-4 \mathcal{A}\mathcal{C}}}{2\mathcal{A}}\,,\label{eq:fieldeq}
\end{align}
where 
\begin{align*}
    \mathcal{A}\equiv \frac{c_3}{M_p^3}\xi\,,\quad \mathcal{B}\equiv 1-\frac{3c_2}{M_p^2}\xi H^2\,, \quad 
    \mathcal{C}\equiv V_{,s} +12 c_1 \xi_{,s} H^4
    \,.
\end{align*}
In obtaining Eqs.~(\ref{eq:friedman}) and (\ref{eq:fieldeq}), we assumed $\dot{\xi}/(2\xi H)\ll1$ with $\xi(s)\neq 0$. Without any loss of generality, one can rewrite Eq.~(\ref{eq:fieldeq}) as
\begin{align}\label{eq:srfieldeq}
    3H \dot{s} \simeq -V_{,s}\left[ 1+ \delta(s)\right]\,,
\end{align}
where 
\begin{align}\label{eq: delta}
    \delta(s)\equiv \frac{\mathcal{B}\mp \sqrt{\mathcal{B}^2-4 \mathcal{A}\mathcal{C}}}{2\mathcal{A}V_{,{s}}}-1\,.
\end{align} 
The duration of inflation is measured by the so-called number of the $e$-folds, which is defined as $N\equiv \int_{t_i}^{t_e} H dt$, where the variables $t_{i}$ and $t_e$ represent the initial and end times of inflation, respectively. Using Eq.~(\ref{eq:srfieldeq}), we obtain  
\begin{align}\label{eq:Nfold}
    N = \int_{s_i}^{s_e}\frac{H}{\dot{s}}ds
    \simeq \frac{1}{M_p^2}\int_{s_e}^{s_i}\frac{V}{V_{,\phi}}\left( \frac{ds}{d\phi}\right)^2\frac{1}{(1+\delta)}d\phi\,.
\end{align}
where $s_i$ and $s_e$ refer to the scalar field values at the beginning and end of inflation, respectively. Here, Eq.~(\ref{eq:Nfold}) shows that the Gauss-Bonnet contributions in the Einstein frame impact the number of $e$-folds through the $\delta(s)$ function as defined in Eq.~(\ref{eq: delta}).

To reflect the aforementioned slow-roll approximations, it is useful to introduce the following so-called slow-roll parameters~\footnote{Our definitions of the $\epsilon_{1}$ and $\epsilon_2$ are different from those in conventional inflation models, where $\epsilon_1$ and $\epsilon_2$ are often defined with the overall minus signs, i.e., $\epsilon_1\equiv -\dot{H}/H^2$ and $\epsilon_2\equiv -\ddot{s}/(H\dot{s})$.}
\begin{align}\label{eq: srp12}
    \epsilon_1&\equiv \frac{\dot{H}}{H^2}\simeq  
    -\epsilon_V (1+\delta)\,, \nonumber\\ 
    \epsilon_2&\equiv \frac{\ddot{s}}{H\dot{s}}\simeq 
    \left[\epsilon_V-\eta_V-\sqrt{2\epsilon_V}M_p\ln(1+\delta)_{,s}\right](1+\delta)\,,
\end{align}
where 
\begin{align*}
    \epsilon_V \equiv \frac{M_p^2}{2}\left( \frac{V_{,s}}{V}\right)^2\,, \quad \eta_V \equiv M_p^2\frac{V_{,ss}}{V}\,.
\end{align*}
Following Ref.~\cite{Hwang:2005hb}, we also introduce the following additional slow-roll parameters to take the effects of the Gauss-Bonnet contributions into account
\begin{align}\label{eq: srp345}
    \epsilon_3 &\equiv \frac{\dot{E}}{2EH} = \frac{E_{,s}}{2E}\frac{\dot{s}}{H}\,,\nonumber\\
    \epsilon_4&\equiv\frac{Q_a}{4M_p^2HQ_t}\,,\\
    \epsilon_5&\equiv \frac{\dot{Q}_t}{2 Q_t H}=\frac{Q_{t,s}}{2Q_t}\frac{\dot{s}}{H}
    \,,\nonumber
\end{align}
where 
\begin{align*}
    E&\equiv \frac{1}{\dot{s}^2}\left(\dot{s}^2+\frac{3Q_a^2}{2M_p^2Q_t}+Q_c \right)
    \,,
\end{align*}
with 
\begin{align*}
    &Q_a\equiv -4c_1\dot{\xi}H^2+\frac{2c_2}{M_p^2}\xi\dot{s}^2H+\frac{c_3}{M_p^3}\xi\dot{s}^3\,,\nonumber\\
    &Q_b\equiv-8c_1\dot{\xi}H +\frac{c_2}{M_p^2} \xi\dot{s}^2\,, \nonumber\\ 
    &Q_c\equiv -\frac{3c_2}{M_p^2}\xi\dot{s}^2H^2+\frac{2c_3}{M_p^3}\dot{s}^3(\dot{\xi}-3\xi H)\,, \nonumber\\
    &Q_t\equiv1+\frac{Q_b}{2M_p^2}\,.
\end{align*}
For slow-roll inflation to occur successfully in our model, we require these slow-roll parameters to be smaller than unity, i.e., $|\epsilon_{1,2,3,4,5}|\ll1$, during inflation. Then, inflation ends as the condition $|\epsilon_1(s_e)|= 1$ is reached.
The $s_e$ value is also affected by the presence of Gauss-Bonnet contributions. Following the linear perturbation analyses carried out in Ref.~\cite{Hwang:2005hb}, we obtain the spectral indices for scalar and tensor fluctuation modes~\cite{Hwang:2005hb}
\begin{align}\label{eq:nsandnt}
    n_S-1 = 2(2\epsilon_1-\epsilon_2-\epsilon_3)\,, \quad n_T = 2(\epsilon_1-\epsilon_5)\,,
\end{align}
and the tensor-to-scalar ratio
\begin{align}\label{eq:rexpression}
    r=16\left| \frac{1}{Q_t}\left(\frac{c_A}{c_T}\right)^3\left(\epsilon_1 -\frac{2Q_c+Q_d -H Q_e+H^2Q_f}{4M_p^2H^2} \right)\right| 
    \,.
    \end{align}
Here, the squared propagation speeds of the scalar and tensor perturbation modes are given ~\cite{Hwang:2005hb, Odintsov:2020ilr} by
\begin{align}\label{eq: squaredCs}
    c_A^2&\equiv 1+\frac{Q_d+\frac{Q_a}{2M_p^2 Q_t}Q_e+\left(\frac{Q_a}{2M_p^2Q_t}\right)^2Q_f}{\dot{s}^2 + \frac{3Q_a^2}{2M_p^2Q_t}+Q_c}\,,\nonumber\\
    c_T^2 &\equiv 1-\frac{Q_f}{2M_p^2Q_t}\,,
\end{align}
where 
\begin{align*}
   & Q_d\equiv -\frac{2c_2}{M_p^2}\xi\dot{s}^2\dot{H} -\frac{2c_3}{M_p^3}\dot{s}^2(\dot{\xi}\dot{s}+\xi\ddot{s}-\xi\dot{s}H)\,,\quad \nonumber\\
   &Q_e\equiv-16c_1\dot{\xi}\dot{H}+\frac{2c_2}{M_p^2}\dot{s}(\dot{\xi}\dot{s}+2\xi\ddot{s}-2\xi\dot{s}H)-\frac{4c_3}{M_p^3}\xi\dot{s}^3\,,\\ 
   &Q_f\equiv8c_1(\ddot{\xi}-\dot{\xi}H)+\frac{2c_2}{M_p^2}\xi\dot{s}^2\,.\nonumber
\end{align*}
When $\alpha=0$, the GR limit, the $Q_{a,b,c,d,e,f}$ quantities vanish, while $Q_t$ becomes unity. Consequently, $\{c_A, c_T\}\rightarrow1$ and the canonical case is restored. When $\alpha\neq 0$, on the other hand, the propagation speeds deviate from the unity. However, if the $c_A$ is either a negative ($c_A<0$) or superluminal ($c_A>1$), one must worry about the ghost instability~\cite{Hwang:2005hb,Odintsov:2020ilr}. We perform full numerical analyses for the values of the $c_A$ and $c_T$ of our model later in this section. 

Now that we have the key observable quantities, we will conduct numerical analyses in the following using Eqs.~(\ref{eq:nsandnt}) and (\ref{eq:rexpression}) and put constraints on the model parameters. In general, we have three free parameters, including $\alpha$, $\lambda$, and $\sigma$. However, if we adopt the Planck normalization~\cite{DeSimone:2008ei} for $\lambda/\sigma^2\sim\mathcal{O}(10^{-9})$ in our numerical study, our model becomes a one-parameter model effectively. This adaptation and the absence of the Gauss-Bonnet contributions, i.e., $\alpha=0$, allow us to recover well-known results of conventional Higgs inflation in the Einstein frame. 

\begin{figure}[tbp]
    \centering
    {\includegraphics[width=0.45\textwidth]{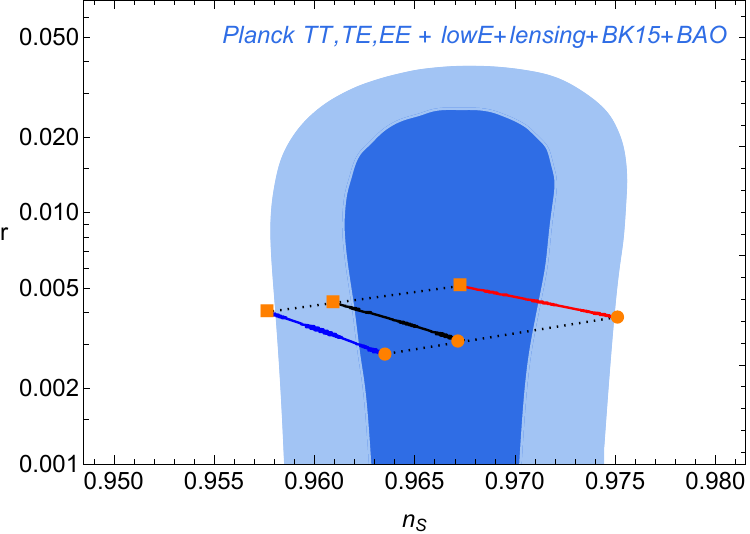}}
    \caption{Numerical plot of $n_S$ vs. $r$ (top) and their number of $e$--fold dependence (bottom) from Eqs.~(\ref{eq:nsandnt})--(\ref{eq:rexpression}). The two ends of each solid line denote $N_\ast=50$(squares) and $N_\ast=60$ (disks). The solid black line ($\alpha=0$) indicates the absence of the Gauss-Bonnet contribution. The model parameter $\alpha$ varies along the black-dotted lines between $-1.4\times 10^4 \text{(red)}\leq \alpha \leq 8\times 10^{3}\text{(blue)}$.}
    \label{fig:01}
\end{figure}
Figure~\ref{fig:01} presents the theoretical predictions of our model in the $n_S$ vs. $r$ plane, along with the observational data. The background dark- and light-blue contours represent the $1\sigma$ and $2\sigma$ confidence level (C.L.) of the $\textit{Planck\, TT, TE, EE+lowE+lensing+BK15+BAO}$ data, respectively. At the same time, the blue, black, and red lines show theoretical predictions of our model for $\alpha=8\times10^3$, $\alpha=0$, and $\alpha=-1.4\times10^4$, respectively. The orange squares and disks denote the $N_\ast=50$ and $N_\ast=60$ $e$-folds, respectively. 
The solid black line in the figure indicates the absence of the Gauss-Bonnet contributions ($\alpha=0$), and in this case, we recover theoretical predictions of Higgs inflation in the GR case. 

In the presence of the Gauss-Bonnet contributions ($\alpha\neq0$), the theoretical predictions of our model shift along the dotted-black lines. Moreover, both the $n_S$ and $r$ values decrease (increase) for the positive (negative) values of $\alpha$. The preferred parameter range of $\alpha$ is between $-1.4\times 10^4 \leq \alpha \leq 8\times10^3$, along the dotted lines. The small wiggles in the plot manifest the numerical errors that accumulated in solving the background equations of motion in Eqs.~(\ref{eq: bEOMs}); hence, they do not indicate any specific significance.~\footnote{The $n_S$ and $r$ in Eqs.~(\ref{eq:nsandnt}) and (\ref{eq:rexpression}) expressions depend on $\ddot{s}$ through the $\epsilon_2$ parameter as defined in Eq.~(\ref{eq: srp12}). Thus, to estimate the $\ddot{s}$, we numerically solve Eqs.~(\ref{eq: bEOM2}) and (\ref{eq: bEOM3}) for $\{s, \dot{s}, H\}$ with appropriate initial conditions, i.e., $\{s_0,\dot{s}_0, H_0\}=\{5.6M_p,0,\sqrt{V(s_0)/(3M_p^2)}\}$. Then, by using the obtained numerical solutions, we approximate the $\ddot{s}$ as a function of time. As a consequence of our numerical treatment, the numerical error manifests in the approximation of $n_S$ and $r$ in Figure~\ref{fig:01}. The numerical error is actually quite small of degree $\mathcal{O}(10^{-14})$ if taken over the whole range of inflation and is `apparently' noticeable in Figure~\ref{fig:01} due to the plotting range of $e$-fold $N$. We use the number of $e$-folds, which is related to time $t$ via $dN=Hdt$, as a time parameter, and the duration of inflation is counted from the end of inflation.} The magnitude of $\alpha$ being relatively large, i.e., $|\alpha|\gg\mathcal{O}(1)$, means our model requires relatively large values of $|\alpha|$ to give rise to noticeable deviations from the GR because the bare contributions of the terms inside the round brackets in Eq.~(\ref{eq: fullaction}) (without multiplying by $\xi(s)$), in general, are quite small. 
\begin{figure}
    \centering
    \includegraphics[width=0.38\textwidth]{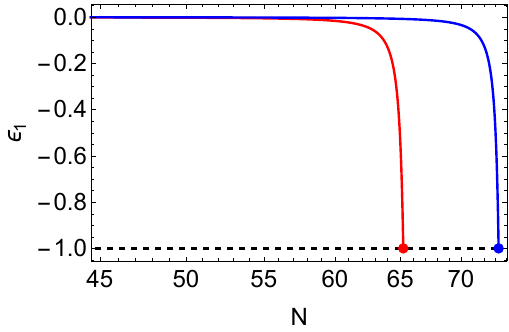}
    \includegraphics[width=0.38\textwidth]{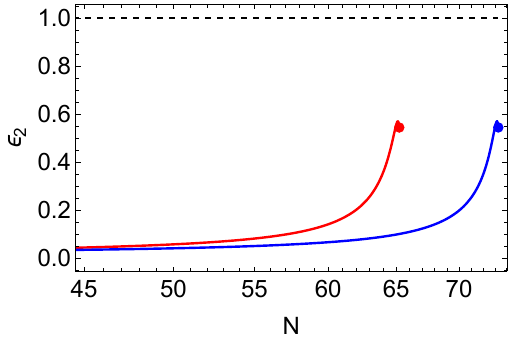}
    \includegraphics[width=0.38\textwidth]{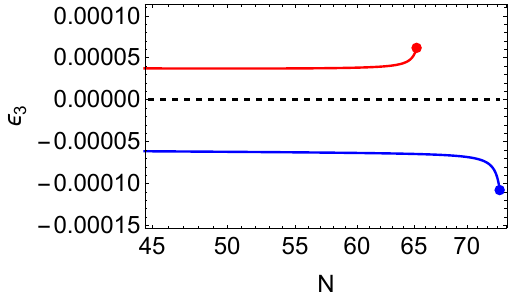}
    \includegraphics[width=0.38\textwidth]{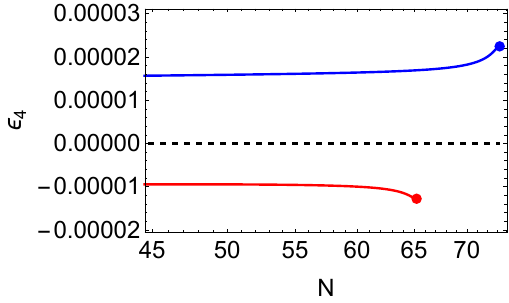}
    \includegraphics[width=0.38\textwidth]{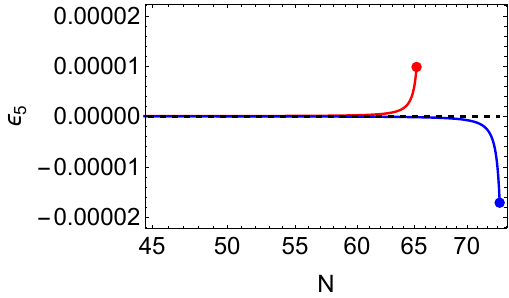}
    \caption{Numerical plot of $\epsilon_{i}(N)$ from Eqs.~(\ref{eq: srp12}) and (\ref{eq: srp345}), where $i=1,2,3,4,5$, for $\alpha=-1.4\times 10^4$(red) and $\alpha=8\times 10^3$(blue). The red and blue disks mark the end time of inflation 
    for each case.}
    \label{fig:02}
\end{figure}

Since our analyses in this section are purely numerical, we need to ensure the smallness of the slow-roll parameters during inflation. Figure~\ref{fig:02} shows the slow-roll parameters ($\epsilon_{i}$ with $i$ ranging from $1$ to $5$) are indeed small during inflation: $|\epsilon_{i}|\lesssim1$ for the values of $\alpha$ that are favored by observational data. The red and blue disks in each sub-figure mark the end time $N_\text{end}$ of inflation, which is determined from $|\epsilon_1(N_\text{end})|=1$. In the absence of Gauss-Bonnet contributions, the $\epsilon_{3,4,5}$ values become zero, the dashed black lines in the figure. For the conventional models of inflation, the tensor power spectrum is called red-tilted (blue-tilted) if the tensor spectral index $n_T$ is negative (positive). From Eq.~(\ref{eq:nsandnt}), the $n_T$ can be negative if the $\epsilon_5>\epsilon_1$. However, Figure~\ref{fig:02} shows for our model that the $\epsilon_1$ parameter is negative during inflation due to our definitions in Eq.~(\ref{eq: srp12}), and it significantly outweighs the $\epsilon_5$ throughout inflation. As a result, the tensor spectral index is negative during the $n_T<0$, and the power spectrum of the tensor fluctuations is, therefore, red-tilted.

The direct detections of GWs from a neutron star merger GW170817~\cite{LIGOScientific:2017vwq}, as well as its associated electromagnetic counterpart GRB170817A~\cite{LIGOScientific:2017zic}, allows us to constrain the GWs propagation speed with remarkable precision: $-3\times10^{-15}\leq c_T/c_\gamma-1\leq 7\times 10^{-16}$ where $c_\gamma$ is the speed of light and we normalize it to $c_\gamma=1$. This bound indicates that the difference in the propagation speed between light and gravitational waves is less than about one part in $10^{15}$. However, the bound corresponds to the late-time universe, where the scalar field value in our model must have reached zero, i.e., $s=0$.  
\begin{figure}[tbp]
    \centering
    \includegraphics[width=0.45\textwidth]{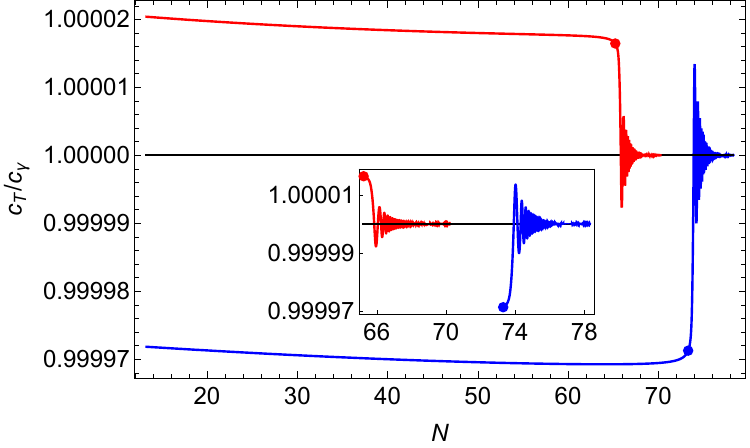}\quad
    \includegraphics[width=0.45\textwidth]{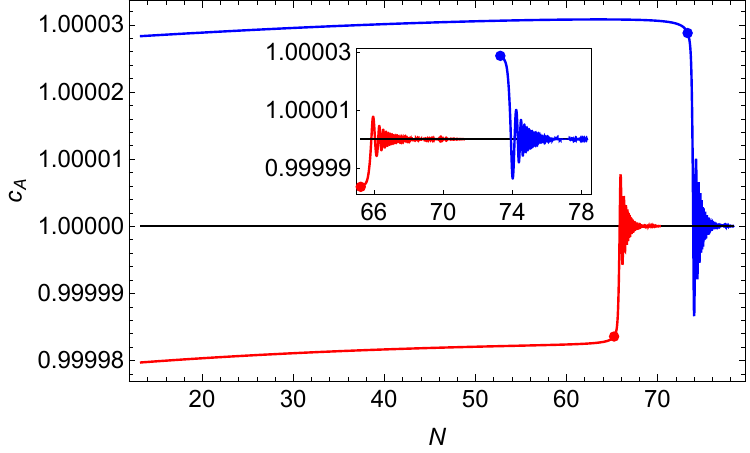}
    \caption{Numerical plot from Eq.~(\ref{eq: squaredCs}) where $c_\gamma(=1)$ is the speed of light. The red and blue lines denote $\alpha=-1.4\times 10^4$ and $\alpha=8\times 10^3$, respectively. The red and blue disks mark the end time of inflation 
    for each case. The horizontal black solid lines at ``$1$'' indicate the GR limit where $c_T=1=c_A$. }
    \label{fig:03}
\end{figure}
When $s\neq0$, which is the case for the early universe, one can expect significant deviations out of this bound induced by the Gauss-Bonnet contributions. Such deviations are subject to future probes. In Figure~\ref{fig:03}, we plot $c_T/c_\gamma$ and $c_A$ as functions of $N$. The red and blue lines denote $\alpha=-1.4\times 10^4$ and $\alpha=8\times 10^3$, respectively, and the ends of inflation for each case are marked with the red and blue disks. The figure, especially the insets, shows that the Gauss-Bonnet contributions gradually decay away and become negligible a few $e$--folds after the end of inflation. As a result, we conclude that Gauss-Bonnet contributions play a significant role during inflation by letting the GWs propagate at a speed different from the speed of light and become negligible over time such that the GW propagation speed converges to that of the speed of light a few $e$-fold after the end of inflation. 

\section{Conclusion}\label{sec: conclusion}
We have investigated Higgs inflation with a Gauss-Bonnet term in the Einstein frame for the model given in Eq.~(\ref{eq: action}). Our model in the Jordan frame has two coupling functions, $\Omega^2(\phi)$ and $\omega(\phi)$, coupled respectively to the Ricci scalar and the Gauss-Bonnet combinations. We assumed these two coupling functions to hold a relation presented in Eq.~(\ref{eq: omegas}) to simplify the delivery of results in the Einstein frame. Our key analytic result of the current work is derived in Eq.~(\ref{eq: fullaction}), where additional interactions, including a non-minimal kinetic coupling between the scalar field and gravity, as well as a derivative self-interaction of the scalar field, emerged in the Einstein frame as a result of a conformal transformation from the Jordan to the Einstein frame. 

From Eq.~(\ref{eq: fullaction}), the background equations of motion are derived in Eq.~(\ref{eq: bEOMs}), and the observable quantities are obtained in Eq.~(\ref{eq:nsandnt}) and (\ref{eq:rexpression}), where we have followed Ref.~\cite{Hwang:2005hb} closely. Although there are three free parameters in the model, including the potential parameter $\lambda$, the non-minimal coupling parameter $\sigma$ between the scalar field and the Ricci scalar, and the coupling parameter $\alpha$ of the Gauss-Bonnet contributions, we showed that our model becomes effectively the one-parameter model if we adopt the Planck normalization for $\lambda/\sigma^2\sim\mathcal{O}(10^{-9})$. The key numerical result of our current work is presented in Figure~\ref{fig:01}, where the theoretical predictions $\{n_S, r\}$ of our model are plotted together with the observational data. 

Without the Gauss-Bonnet contributions, where $\alpha=0$, our result recovers the predictions of Higgs inflation in the GR. Once the Gauss-Bonnet contributions are turned on with $\alpha\neq0$, the $n_S$ and $r$ predictions deviate from the GR case. The $n_S$ and $r$ values decrease (increase) for the positive (negative) $\alpha$ values, as is seen in Figure~\ref{fig:01}. The observational data favor the broad range model parameter, $-1.4\times 10^4\lesssim \alpha \lesssim 8\times 10^4$. 
In Figure~\ref{fig:03}, our analysis reveals that the propagation speed of gravitational waves (GWs) deviates from the speed of light during the inflationary period, influenced by Gauss-Bonnet contributions on the order of a few parts in hundreds of thousands. These Gauss-Bonnet effects gradually dissipate after the inflation, leading the GWs to progressively align with the speed of light.
We have also shown the validity of the slow-roll approximation in Figure~\ref{fig:02} by showing the slow-roll parameters are small, much smaller than unity, during inflation.

In our future research, we plan to relax our assumption made in Eq.~(\ref{eq: omegas}) and explore post-inflationary cosmology and its implications for (p)reheating. It also remains to be determined whether these newly emerged interactions in the Einstein frame can adequately account for the observed late-time accelerating expansion of the universe.

\begin{acknowledgements}
The authors acknowledge that this work was supported by the Basic Science Research Program through the National Research Foundation of Korea (NRF), funded by the Ministry of Education (grant numbers) (NRF-2022R1I1A1A01053784) (GT), (NRF-2021R1A2C1005748) (SK) and by (NRF-2021R1A4A2001897), (NRF-2019R1A2C1089334) (SCP).
\end{acknowledgements}

\appendix
\section{Constant coupling to the Gauss-Bonnet term}\label{sec: appA}
\renewcommand{\theequation}{A.\arabic{equation}}\setcounter{equation}{0}
Let us consider the constant Gauss-Bonnet coupling, i.e., $\omega(\phi)=\text{const.}$ in Eq.~(\ref{eq: action}), and simplify the Einstein frame action in Eq.~(\ref{eq: actionGB1}). For the fourth term in Eq.~(\ref{eq: actionGB1}), we use
\begin{align}\label{eq: identity}
    &\nabla_a \left[\Omega^{-2}\left(\nabla_b\nabla^b\Omega \nabla^a\Omega -\frac{1}{2}\nabla^a(\nabla\Omega)^2 \right)\right] \nonumber\\
    &\quad = \Omega^{-2}\left[(\nabla_a\nabla^a\Omega)^2 - (\nabla_a\nabla_b\Omega)^2\right]- R^{ab} \Omega^{-2} \nabla_a\Omega\nabla_b\Omega \nonumber\\ &\qquad -2\Omega^{-3} (\nabla\Omega)^2\nabla_b\nabla^b\Omega 
    +2\Omega^{-3} \nabla_a\Omega\nabla_b\Omega\nabla^a\nabla^b\Omega\,,
\end{align}
and integration by parts to obtain
\begin{align}\label{eq: term1}
    &8 \omega \int d^4x \sqrt{-g}\,\Omega^{-2} \left(\nabla_a\nabla^a\Omega\nabla_b\nabla^b\Omega-\nabla_b\nabla_a\Omega\nabla^b\nabla^a\Omega \right)\nonumber\\
    &= \omega \int d^4x \sqrt{-g} \left[2 R^{ab} \nabla_a \ln \Omega^2 \nabla_b\ln \Omega^2 +4(\Omega^{-1}\nabla_b\nabla^b\Omega) \right. \nonumber\\
    &\left. \quad \times(\nabla \ln \Omega^2)^2-4(\Omega^{-1}\nabla^a\nabla^b\Omega)(\nabla_a \ln \Omega^2 \nabla_b\ln \Omega^2)\right]\,,
\end{align}
where the following relations are used
\begin{align}
    &\nabla_a \ln \Omega = \Omega^{-1}\nabla_a\Omega  \,,\nonumber\\
    &\Omega^{-1} \nabla^a\nabla_a\Omega = \nabla^a\nabla_a \ln \Omega+ \nabla^a\ln\Omega\nabla_a\ln\Omega \nonumber \\
    &\qquad\qquad= \frac{1}{2}\nabla^a\nabla_a \ln \Omega^2 +\frac{1}{4} \nabla^a\ln\Omega^2\nabla_a\ln\Omega^2\,.\label{eq: useful}
\end{align}
The third term in Eq.~(\ref{eq: actionGB1}) can also be rewritten as 
\begin{align}\label{eq: term2}
    &-4 \omega \int d^4x\sqrt{-g}R \Omega^{-2} \nabla_a \Omega \nabla^a \Omega \nonumber \\
    &\qquad = -\omega\int d^4x\sqrt{-g} g^{ab}R \nabla_a \ln \Omega^2 \nabla_b \ln \Omega^2\,.
\end{align}
Then, the first term on the right-hand side of equality in Eq.~(\ref{eq: term1}) is combined with Eq.~(\ref{eq: term2}) to give
\begin{align}
    2\omega \int d^4x\sqrt{-g} G^{ab} \nabla_a\ln \Omega^2 \nabla_b\ln \Omega^2,
\end{align}
which is then canceled with the second term in Eq.~(\ref{eq: actionGB1}). ~\footnote{The integration by parts of the second term in Eq.~(\ref{eq: actionGB1})
\begin{align}
    -2\omega\int d^4x\sqrt{-g}G_{ab}\nabla^a\ln\Omega^2\nabla^b\ln \Omega^2\,.\nonumber
\end{align}
} 
Thus, the remaining terms in Eq.~(\ref{eq: actionGB1}) reads 
\begin{align}\label{eq: App6}
    S&=\int d^4x\sqrt{-g}\, \omega \left[R_{GB}^2 
    -2(\Omega^{-1}\nabla_b\nabla^b\Omega)(\nabla \ln \Omega^2)^2 \right. \nonumber \\ 
    &\quad-4(\Omega^{-1}\nabla^a\nabla^b\Omega)(\nabla_a \ln \Omega^2 \nabla_b\ln \Omega^2) \nonumber\\
    &\qquad\left. +\frac{3}{2}\left(\nabla_a\ln \Omega^2 \nabla^a\ln\Omega^2\right)^2 \right]\,.
\end{align}
Let us rewrite Eq.~(\ref{eq: App6}) once again using Eq.~(\ref{eq: useful})
\begin{align}
    S&=\int d^4x\sqrt{-g}\, \omega \left[R_{GB}^2 -(\nabla_b\nabla^b\ln \Omega^2) \right.\nonumber\\
    &\quad \times\left(\nabla_a\ln \Omega^2 \nabla^a\ln\Omega^2\right)-2(\nabla^b\nabla^a \ln\Omega^2)\nonumber\\
    &\qquad \times(\nabla_a \ln \Omega^2 \nabla_b\ln \Omega^2)-(\nabla_a \ln \Omega^2 \nabla_b\ln \Omega^2)^2 \nonumber\\
    &\left.\qquad\quad+\left(\nabla_a\ln \Omega^2 \nabla^a\ln\Omega^2\right)^2\right]\,. 
\end{align} 
The second and third terms are canceled after integration by parts. Thus, we obtain 
\begin{align}
    S &= \int d^4x\sqrt{-g}\, \omega \left[R_{GB}^2 -\frac{4}{9}(\nabla_a s \nabla_b s )^2+\frac{4}{9} \left(\nabla_a s \nabla^a s \right)^2\right]\nonumber\\
    &= \int d^4x\sqrt{-g}\, \omega R_{GB}^2\,,
\end{align}
where $s\equiv\sqrt{3/2}\ln \Omega^2$. It is well-known in the literature that the Gauss-Bonnet term is topological in $4$--dimensions if the Gauss-Bonnet coupling is a constant. Thus, for the $\omega=\text{const.}$ case, we conclude that no dynamical contributions emerge from the Gauss-Bonnet term in the Jordan and Einstein frames.

\section{Power-law coupling to Gauss-Bonnet term}\label{sec: appB}
\renewcommand{\theequation}{B.\arabic{equation}}\setcounter{equation}{0}
Let us now assume the coupling functions in Eq.~(\ref{eq: action}) hold a more general relation as $\omega=\alpha \Omega^p$. When $p=2$, the action in the Jordan frame reads 
\begin{align}\label{eq: AppB1}
    S^J=\int d^4x \sqrt{-g^J} \left[ \Omega^2(\phi) \left(
    \frac{M_p^2}{2} R^J +\alpha R_{GB}^2 \right) \right.\nonumber\\
    \left.- \frac{1}{2}g^J_{ab}\nabla^a\phi\nabla^b\phi -V(\phi)\right].\,
\end{align}
The Einstein frame action is presented in Eq.~(\ref{eq: actionGB3}). 

sFor $\omega=\alpha \Omega^p$ with arbitrary power of $p$, the third term of Eq.~(\ref{eq: actionGB3}) can be written as 
\begin{align}
    8\alpha (p-2) \Omega^{p-3}\nabla_a \Omega \left(\nabla^a\Omega\nabla_b\nabla^b\Omega -\nabla_b\Omega\nabla^a\nabla^b\Omega\right)\,.
\end{align}
Consequently, Eq.~(\ref{eq: actionGB3}) becomes
\begin{align}\label{eq: actionGBp1}
    S&=\int d^4x\sqrt{-g}\alpha \Omega^p\left[ R_{GB}^2 + 8p\Omega^{-2}G_{ab}\nabla^a\Omega\nabla^b\Omega \right.\nonumber\\ 
    &\quad - 8(p+1) \Omega^{-3}\nabla_a \Omega\nabla^a\Omega\nabla_b\nabla^b\Omega \nonumber\\  
    &\qquad+8(p-2) \Omega^{-3}\nabla_a \Omega\nabla_b\Omega\nabla^a\nabla^b\Omega \nonumber\\
    &\left.\qquad\quad+24\Omega^{-4}\left( \nabla_a\Omega \nabla^a\Omega\right)^2\right]\,.
\end{align}
Applying Eqs.~(\ref{eq: useful}) to the last term in the first line, we obtain
\begin{align}\label{eq: actionGBp2}
    S&=\int d^4x\sqrt{-g}\alpha  e^{\frac{p}{2}\sqrt{\frac{2}{3}}\frac{s}{M_p}}\left[R_{GB}^2 +\frac{4p}{3M_p^2}G_{ab}\nabla^a s\nabla^b s \right.\nonumber\\
    &\quad- \frac{p+1}{3 M_p^3}\sqrt{\frac{8}{3}}\nabla_b\nabla^b s \nabla_a s \nabla^a s \nonumber\\
    &\left.\qquad+\frac{p-2}{3M_p^3}\sqrt{\frac{8}{3}}\nabla_a s \nabla_b s \nabla^a\nabla^b s\right]\,,
    \end{align}
where $(s/M_p)\equiv\sqrt{3/2}\ln\Omega^2$. The last term vanishes for $p=2$ and we get Eq.~(\ref{eq: actionGB5}). As a result of conformal transformation from the Jordan frame to the Einstein frame, we notice the emergence of new interactions such as the kinetic coupling between the scalar field and gravity and the derivative self-interactions of the scalar field. These interactions certainly would contribute both to the background and the perturbation dynamics.

\end{document}